\documentclass[preprint,journal]{vgtc}       

\onlineid{0}

\usepackage{wrapfig}

\ieeedoi{10.1109/TVCG.2023.3326576}

\vgtccategory{Research}

\vgtcpapertype{application/design study}

\title{Reclaiming the Horizon: Novel Visualization Designs \\ for Time-Series Data with Large Value Ranges}

\author{%
  \authororcid{Daniel\ Braun}{0000-0002-8824-7184},
  \authororcid{Rita\ Borgo}{0000-0003-2875-6793}, 
  \authororcid{Max\ Sondag}{0000-0003-3309-638X}, and 
  \authororcid{Tatiana\ von\ Landesberger}{0000-0002-5279-1444}
}

\authorfooter{
  \item
  	Daniel Braun, Max Sondag, and Tatiana von Landesberger are with University of Cologne.
  	E-mail: \{braun\,$|$\,sondag\,$|$\,landesberger\}@cs.uni-koeln.de.
  \item
  	Rita Borgo is with King's College London.
  	E-mail: rita.borgo@kcl.ac.uk.
}

\abstract{%
 We introduce two novel visualization designs to support practitioners in performing identification and discrimination tasks on large value ranges (i.e., several orders of magnitude) in time-series data: (1) The \textit{order of magnitude horizon} graph, which extends the classic horizon graph; and (2) the \textit{order of magnitude line} chart, which adapts the log-line chart. These new visualization designs visualize large value ranges by explicitly splitting the mantissa $m$ and exponent $e$ of a value $v=m \cdot 10^e$.
 We evaluate our novel designs against the most relevant state-of-the-art visualizations in an empirical user study. It focuses on four main tasks commonly employed in the analysis of time-series and large value ranges visualization: identification, discrimination, estimation, and trend detection. For each task we analyze error, confidence, and response time. The new \textit{order of magnitude horizon} graph performs better or equal to all other designs in identification, discrimination, and estimation tasks. Only for trend detection tasks, the more traditional horizon graphs reported better performance. Our results are domain-independent, only requiring time-series data with large value ranges.
}


\teaser{
  \centering
     \begin{subfigure}[tb]{0.32\textwidth}
         \centering
         \includegraphics[width=\textwidth]{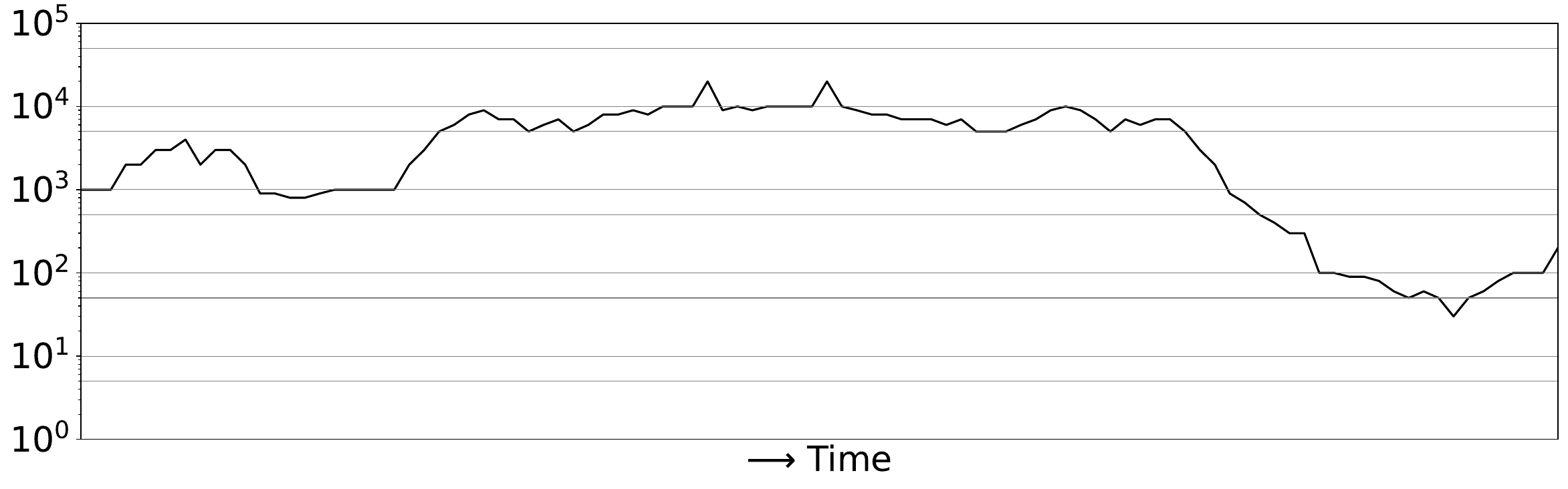}
         \caption{Log-line chart}
         \label{fig:Log}
     \end{subfigure}
     \hfill
     \begin{subfigure}[tb]{0.32\textwidth}
         \centering
         \includegraphics[width=\textwidth]{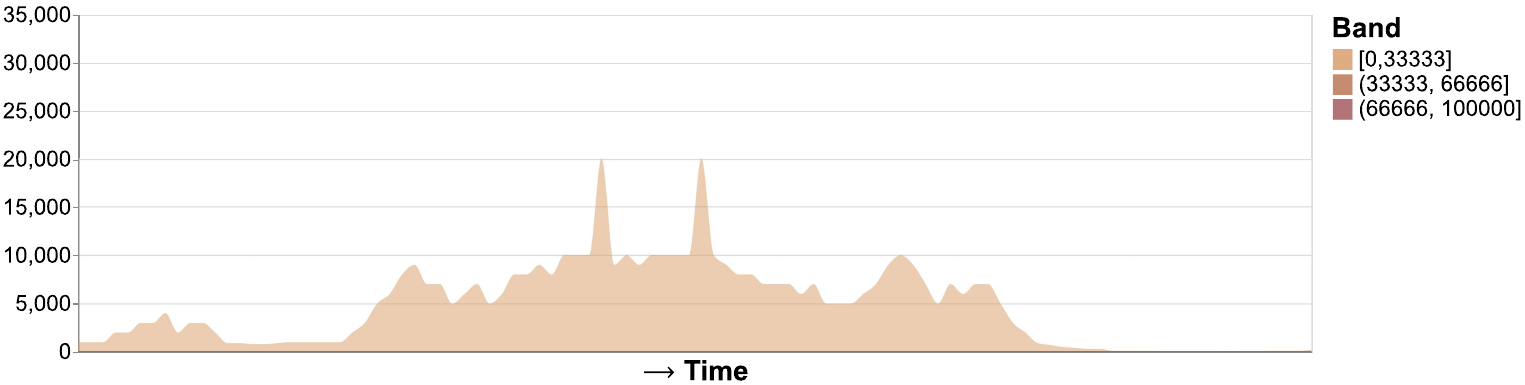}
         \caption{Classic horizon graph~\cite{Few.2008, Reijner.2008}}
         \label{fig:Horizon}
     \end{subfigure}
     \hfill
     \begin{subfigure}[tb]{0.32\textwidth}
         \centering
         \includegraphics[width=\textwidth]{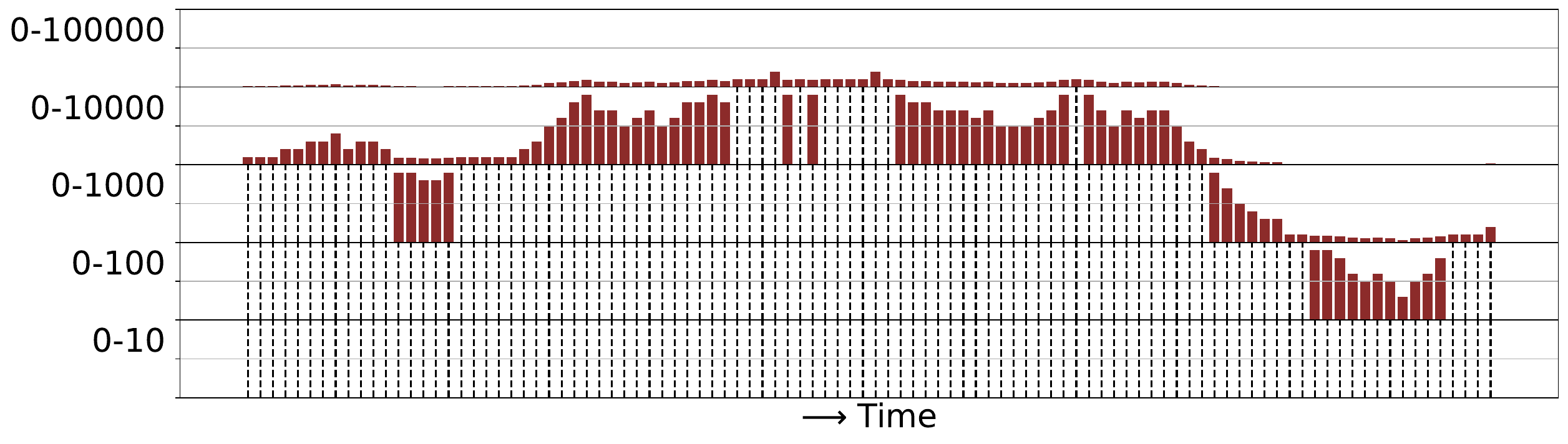}
         \caption{Scale-stack bar chart (SSB)~\cite{Hlawatsch.2013}}
         \label{fig:SSB}
     \end{subfigure}
     
     \begin{subfigure}[tb]{0.49\textwidth}
         \centering
         \includegraphics[width=\textwidth]{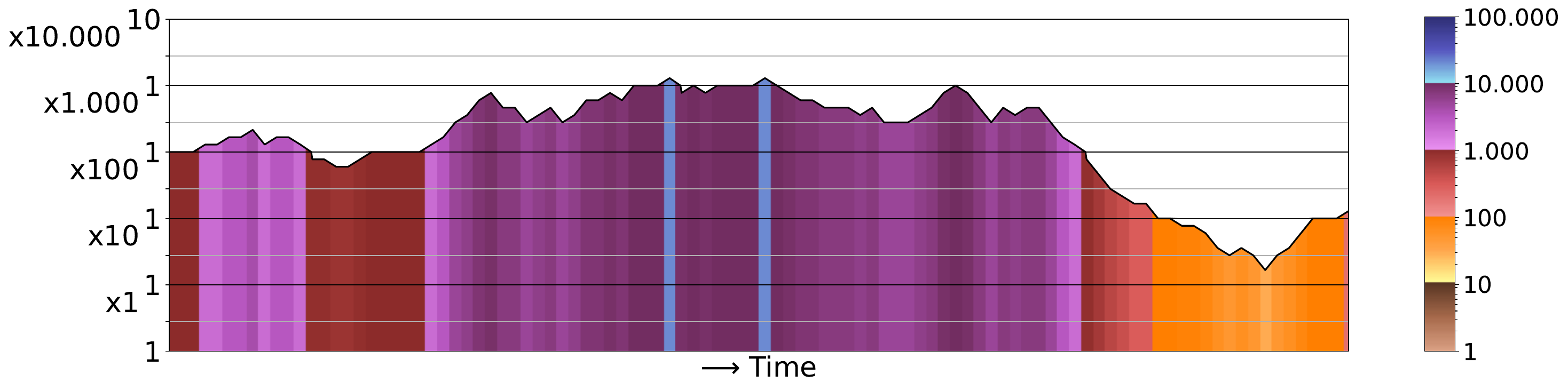}
         \caption{Order of magnitude line chart (OML)}
         \label{fig:OML}
     \end{subfigure}
     \hfill
     \begin{subfigure}[tb]{0.49\textwidth}
         \centering
         \includegraphics[width=\textwidth]{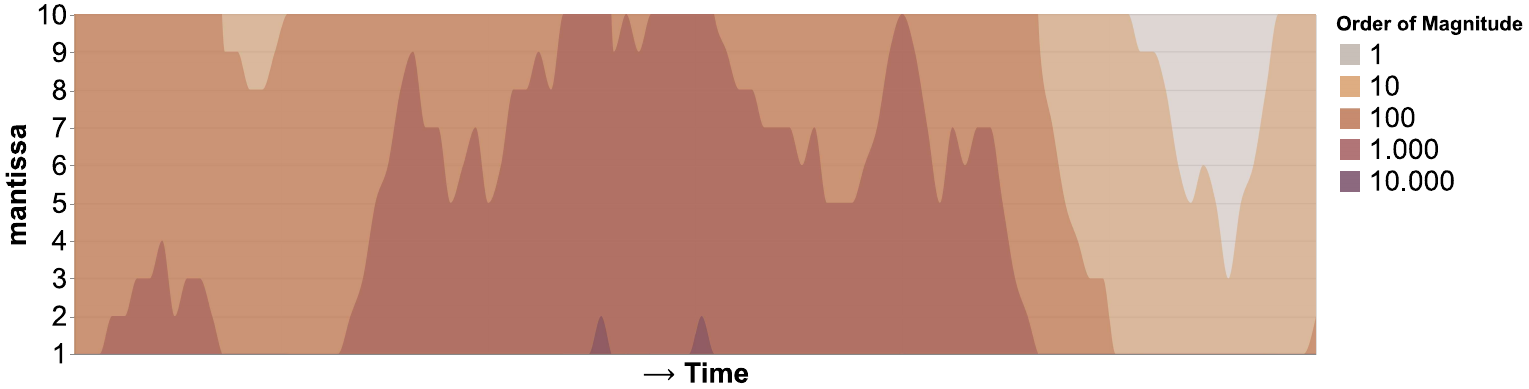}
         \caption{Order of magnitude horizon graph (OMH)}
         \label{fig:OMH}
     \end{subfigure}%
        \subfigsCaption{Exemplar time-series data with large value ranges encoded using three standard visualization types log-line chart (a), horizon graph (b), scale-stack bar chart (c), and our two new designs \textit{order of magnitude line} (d) and \textit{order of magnitude horizon} (e).}
        \label{fig:teaser}
}

\graphicspath{{figs/}{figures/}{pictures/}{images/}{./}} 

\usepackage{tabu}                      
\usepackage{booktabs}                  
\usepackage{lipsum}                    
\usepackage{mwe}                       

\usepackage{mathptmx}                  
\usepackage{newtxmath}

\usepackage{marginnote}

\begin{document}


\firstsection{Introduction}

\maketitle

\begin{figure}[htbp]
 \centering
 \includegraphics[width=0.85\columnwidth]{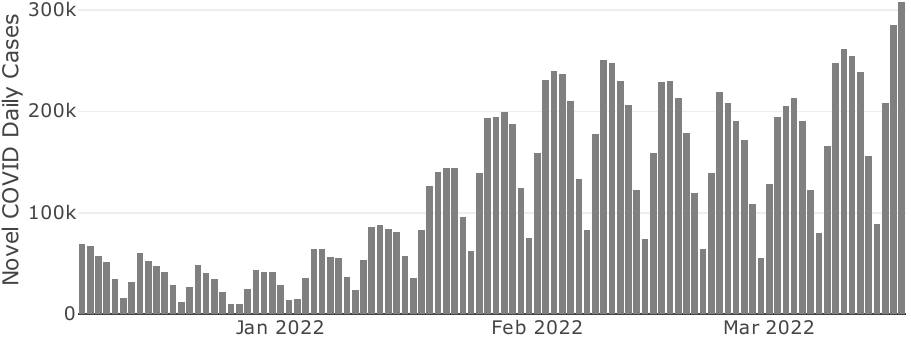}
 \caption{Number of daily new COVID-19 cases in Germany.}
 \label{fig:covid}
\end{figure}

Large value ranges (i.e., several orders of magnitude) in time-series data) are common to a wide range of application domains.
These large variations in orders of magnitude can occur in a relatively short amount of time in a variety of application domains. Examples include medicine with pandemic outbreaks (e.g., COVID-19 cases in Germany, which range from 10,000 to 300,000 per day within 100 days in \cref{fig:covid}), meteorology with measurements of storms (e.g., the ice water content in clouds, which range from $10^{-7}$ to $10^{-3} kg/m^3$ in hours~\cite{Braun.2022}), or finance with stock markets (e.g., bitcoin cryptocurrency with price changes of more than 40,000 dollars within a year~\cite{Bitcoin.2022}). 

Research on visualization of short time-series has proposed techniques to improve accuracy and completion time of low-level tasks compared to standard line charts~\cite{Aigner.2008, Fang.2020}. However, these visualizations fail in visualizing large orders of magnitude.
\Cref{fig:Horizon} shows one such technique: a horizon graph~\cite{Few.2008, Reijner.2008}, which divides and layers the line chart. Here, small changes in values are difficult to detect across all orders of magnitudes, and especially difficult is reading and comparing values at lower magnitudes. In \cref{fig:covid} we see another example of this. At the same time, research on visualizing large value ranges focuses mainly on uni-variate data without time~\cite{Hlawatsch.2013, Hohn.2020, Borgo.2014}. This paper combines these two areas, visualizing large value ranges in time-series data.

We present two novel visualization designs that improve two standard visualization techniques for time-series data -- log-line charts and horizon graphs -- to meet the challenges of magnitude variations. First, the \textit{order of magnitude line (OML)} chart (\cref{fig:OML}) adapts the log-line chart by using linear mapping within each order of magnitude.
In addition, the OMC color scale~\cite{Braun.2022} -- a color scale designed for large value ranges -- is used to support the perception of magnitude variation and value identification. The second proposed design -- the \textit{order of magnitude horizon (OMH)} graph (\cref{fig:OMH}) -- adjusts the standard horizon graph by using a separate representation of the mantissa and exponent of each value. Every order of magnitude is represented by a separate band, while the mantissa is mapped to a linear scale on the y-axis for easier identification of individual values. In this paper, we assume that numbers are encoded in base 10, but other bases could be supported.

To evaluate our new designs, we conduct a user study that compares them with the standard visualization techniques for time-series -- log-line charts (\cref{fig:Log}) and horizon graphs -- as well as one state-of-the-art approach for large value ranges for time-independent data -- the scale-stack bar chart~\cite{Hlawatsch.2013}. Our study focuses on low-level visualization tasks from taxonomies of both time-series~\cite{Aigner.2008, Aigner.2011} and large value ranges~\cite{Borgo.2014, Hohn.2020, Braun.2022, Hlawatsch.2013}. The results show that OMH outperforms their currently used counterparts in tasks of "identification" (i.e., read a marked value), "discrimination" (i.e., compare two marked values in one visualization), and "estimation" (i.e., determine the difference of two marked values). OMH has both the lowest error rate and the highest confidence of all designs in these three tasks, while the time taken by the user is similar to the others. For OML, we only find a significant effect on confidence when compared to horizon graphs. The standard horizon graph on "detect trend" (i.e., identify the trend of the visualized data), tasks proves to be better than all designs tested except OMH, likely due to the linear scaled y-axis as opposed to logarithmic scaling.

The primary \textbf{contributions} of our work are:
\begin{itemize}[noitemsep]
    \item We introduce \textit{two novel visualization designs}, extending existing time-series designs to meet the requirements of data with large value ranges.
    \item We \textit{empirically evaluate} our new designs with state-of-the-art designs in a user study with 90 participants.
    \item The \textit{novel designs are domain-independent} and can hence be used on a variety of application domains (e.g., meteorology, finance, healthcare) to better present temporal data with large value ranges.
\end{itemize}

After an overview of current methods and evaluations in \cref{sec:related}, we introduce our two new designs in \cref{sec:designs}. The design of our user study is presented in \cref{sec:evaluation} and its analysis and results are described in \cref{sec:results}. In addition to a discussion of the findings in \cref{sec:discussion}, we also point out limitations and future work in \cref{sec:limitations}.

\section{Related Work}
\label{sec:related}

We present the latest solutions and evaluations for displaying time-series as well as for visualizing data with varying orders of magnitude.

\paragraph{Visualization of Time-Series}
The visualization of time-series data has been widely investigated using a variety of different visual designs~\cite{Aigner.2008, Fang.2020, Aigner.2011, Adnan.2016}. The most popular way to present continuous coherence in time-series is the line chart~\cite{Albers.2014, Correll.2012}. 
Over time, several novel visualization techniques have been designed and evaluated in relation to the traditional line graph. The combination of quantitative data with qualitative abstraction (e.g., a composite representation with color encoding of qualitative data and spatial position encoding of the quantitative data) has been studied by Aigner et al.~\cite{Aigner.2012} and Federico et al.~\cite{Federico.2014}. We adapt the idea of adding qualitative coloring based on a division of the value range to a line graph in our \textit{order of magnitude line} design. Meaningful color coding is used successfully in other time-series visualizations, such as ThemeRivers~\cite{Havre.2000}, to show thematic changes over time, and RankExplorer to visualize ranking changes over time~\cite{Shi.2012}.

In recent years, the horizon graph has become increasingly popular. The final design was presented for the first time by Reijner~\cite{Reijner.2008} and Few~\cite{Few.2008}, while the principle of a two-tone pseudo coloring was already used some years before by Saito et al.~\cite{Saito.2005}. In a horizon graph, the value range is divided into several bands of the same size, which are then superimposed onto each other to create a layered form. Color (hue and saturation) is used to distinguish the individual bands. 
Heer et al.~\cite{Heer.2009} compared the horizon graph with line graphs and investigated the appropriate number of bands. They found that horizon graphs improve estimation accuracy and using more than three bands increases the error and time. Therefore, we use three bands for the standard horizon graph in our evaluation. Jabbari et al.~\cite{Jabbari.2018, Jabbari.2018b} compare the horizon graph to alternative composite visual mappings, such as a hue-saturation mapping which is very similar to warming stripes~\cite{Hawkins.2018}. They found that the horizon graph performed best in terms of discrimination and estimation errors, and was only slightly slower than the proposed alternative mappings. For this reason, we have decided not to evaluate a warming stripes variant for large value ranges in our study. 
Gogolou et al.~\cite{Gogolou.2018} explore the aspect in comparing similarity perception between line charts, horizon graphs, and colorfields. The horizon graph promotes local variations the most, while the other two have advantages for amplitude and y-offset scaling.

One of the advantages of the horizon graph is the space saved by layering the different bands, which makes it especially efficient for multiple time-series. 
Although we do not focus on multiple time-series in this paper, the results of Javed et al.~\cite{Javed.2010} for horizon graphs in that specific case are interesting. They show faster perception in discrimination tasks for horizon graphs than for simple line graphs and a similar correctness rate compared to the other evaluated designs for every task. The study of Perin et al.~\cite{Perin.2013} supports these results. Here, multiple horizon graphs outperform multiple line charts in each task, with even better results for an interactive version of the horizon graphs. These studies indicate that extending our OMH to multiple time-series could prove promising in future work.

\paragraph{Visualization of Large Value Ranges}
A common way to display time-series with order of magnitude variations is to use a line graph with logarithmic scaling~\cite{Sevi.2020, Romano.2020, Kubina.2022}. However, most of the previous research on large value ranges has been done for time-independent data.

Hlawatsch et al.~\cite{Hlawatsch.2013} were the first to develop a new visualization technique for displaying large-value ranges and adapted the classic bar chart. Their \textit{scale-stack bar chart (SSB)} (\cref{fig:SSB}) represents each value at multiple stacked scales with increasing value ranges. Each scale starts at zero and maps the numbers linearly. The results of a user study showed improvements of their approach compared to linear and logarithmic bar charts for quantitative comparisons.

Borgo et al.~\cite{Borgo.2014} use the separate representation of exponent and mantissa by showing both parts of each value as two overlapping bars on one linear scale in a classic bar chart, and call them \textit{order of magnitude markers}. Their evaluation confirms that designs that consider large value ranges increase the accuracy for evaluation tasks.

By encoding the orders of magnitude using the color and width of classic bar charts, Höhn et al.~\cite{Hohn.2020} adapted the previous method. They compare their \textit{width-scale bar charts} to the two other approaches in a quantitative study. The results showed that their design works best for interpretation tasks. However, for estimation, discrimination, and trend tasks, the SSB resulted in the highest accuracy of the three approaches. Therefore, we decided to include the SSB in our evaluation.

Braun et al.~\cite{Braun.2022} showed that color can be used to support the perception of changes in the orders of magnitude. Their \textit{order of magnitude color (OMC)} colormap encodes a value $v=m \cdot 10^e$ by mapping the exponents $e$ to the hue and the mantissa $m$ to a sequential color scale of that hue. We use this color scheme in our OML design.

\section{Visualization Designs}
\label{sec:designs}

The presented visualization designs combine methods for time-series data as well as methods for data with large value ranges. We designed our visualizations as static, non-interactive visualizations. While interactivity could help in understanding charts with multiple series~\cite{Perin.2013, Fang.2020} or long time-series~\cite{Aigner.2008, Aigner.2011}, we focus on short and singular time-series here. Moreover, the static views can be used in printed media.

The "default" way to visualize large value ranges, is to use a logarithmic scale. However, reading values from a logarithmic scale is not intuitive~\cite{Kubina.2022}.
This is evidenced by the focus of previous techniques for large value ranges on perceptually linear scaling for both the mantissa $m$ and the exponent $e$ of a value $v=m \cdot 10^e$~\cite{Hlawatsch.2013, Borgo.2014, Hohn.2020}. Our approach reflects techniques for large value ranges by splitting the mantissa $m$ from the  exponent $e$ and treating them differently. Values within the same exponent $e$ are visualized continuously, while there is a jump between values with a different exponent. This continuity is an advantage over the SSB approach, where a coherent reading of the data is interrupted by the need to repeatedly restart the scales at $v=0$.

\subsection{Order of Magnitude Horizon}
\label{sec:OMH}

\begin{figure}
\includegraphics{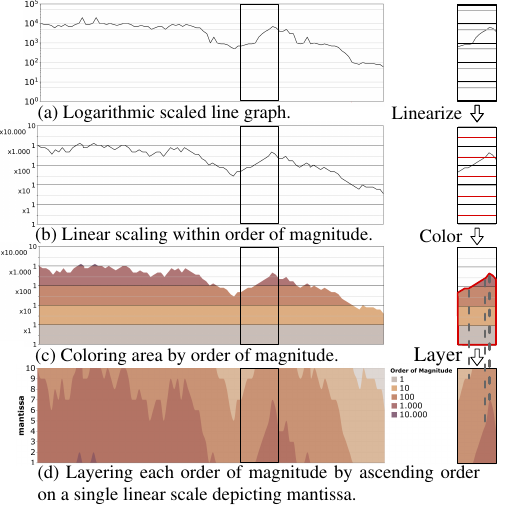}
 \caption{\textit{Order of magnitude horizon} graph: Step-by-step construction.}
  \label{fig:OMHCreationFig}
\end{figure}

The \textit{order of magnitude horizon (OMH)} graph is an adaptation of the standard horizon graph to meet the requirements of data with large value ranges.
\cref{fig:OMHCreationFig} illustrates the construction of OMH. Unlike the standard horizon graph, the starting point is a log-line graph (\cref{fig:OMHCreationFig}a). After scaling the y-axis linearly within each order of magnitude (\cref{fig:OMHCreationFig}b), the graph is split into uniformly-sized bands. Each band represents a different order of magnitude, and hence the number of bands depends on the number of orders of magnitude. We assume that numbers are encoded in base 10, but other bases could be supported. The colors of the bands are based on the hue gradient of the OMC color scale\cite{Braun.2022}: The larger the magnitude of the band, the more saturated its color (\cref{fig:OMHCreationFig}c).
Finally, the different bands are layered on top of each other, sorted in ascending order (\cref{fig:OMHCreationFig}d).
By using two different visual variables, vertical position for the mantissa $m$, and saturation for the exponent $e$, both $m$ and $e$ can be directly compared in our design. 
Even though not evaluated, it should be possible to use OMH for data with both positive and negative values with a diverging color scale.

\subsection{Order of Magnitude Line}

\begin{figure}
    \centering
    \includegraphics{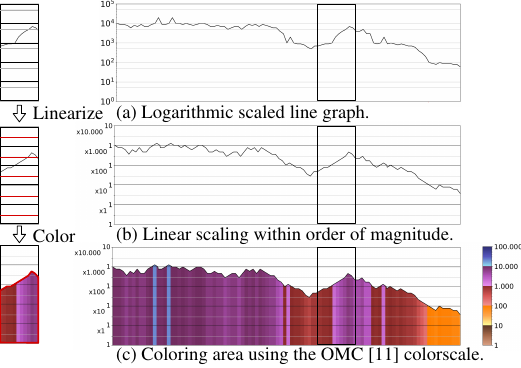}
    \caption{\textit{Order of magnitude line} graph: Step-by-step construction.}
    \label{fig:OMLCreation}
\end{figure}

The order of magnitude line (OML) graph is an adaption of the log-line graph.
Starting off with a log-line graph (\cref{fig:OMLCreation}a), the y-axis is scaled linear within each order of magnitude (\cref{fig:OMLCreation}b) to facilitate easier reading of the values. In addition, the area below the line in the graph is colored with the OMC~\cite{Braun.2022} color scale (\cref{fig:OMLCreation}c) to further improve the perception of the values. Hence, the values are double encoded by the visual variables vertical position and color. In the OMC color scale, each exponent $e$ is given a different hue, with a sequential color scale for the mantissa $m$. This change in hue between the different exponents $e$ reduces the probability of a magnitude error.

\subsection{Design Novelty}
Compared to traditional methods, both OML and OMH designs propose a novel approach with respect to spatial organization of the data. Both designs enforce continuity in the visual layout. This is lost in traditional approaches dealing with visualization of large magnitude values, where priority in the design is given to the enhancement of value variations. Our designs instead intend to favour perceptual principles, leveraging Gestalt~\cite{Wertheimer.1923} principles of grouping through proximity, similarity, continuity and common faith.

\section{Evaluation}
\label{sec:evaluation}

To evaluate our two novel approaches, we compare them to their standard counterparts (\cref{fig:teaser}) -- log-line chart and horizon graph for time-series data as well as SSB charts for data with large value ranges -- in an empirical user study to test the following hypotheses:
\begin{itemize}[noitemsep]
    \item \textbf{H1:} \textit{Our new designs reduce error rates in value identification, discrimination, estimation, and trend detection tasks on data with large value ranges compared to their standard counterparts.} 
    
     As our designs are explicitly designed based on research for large value ranges, which have been proven to improve performance on these tasks~\cite{Braun.2022, Hlawatsch.2013, Hohn.2020, Borgo.2014}, we expect this improvement to translate towards time-series with large value ranges.
     
    \item \textbf{H2:} \textit{Our new designs increase participants' confidence in their answers compared to their standard counterparts.}
    
    We expect that the linear scaling of OMH and OML makes them easier to read. Furthermore, participants have multiple options to confirm their answers due to the double encoding by color.
    
    \item \textbf{H3:} \textit{Our new designs have higher response times than the standard log-line chart and comparable response times to the others.}
    
    Participants should be familiar with standard log-line charts. Due to the novelty, we expect that they will need more time to interpret our designs.
    
\end{itemize}

\subsection{Stimuli Design}
Beyond the question of which visualization techniques to evaluate, there are additional ways to improve charts to allow better interpretation of presented data. Throughout the stimuli design process we followed state-of-the-art visualization guidelines~\cite{Ware.2019, Munzner.2015, Diehl.2018}.

Considering which visual elements to include or omit in the stimuli designs quickly leads to the question of "chartjunk"~\cite{Few.2011}. Research on this topic shows that visual embellishments that are not essential to understanding the data can help to transmit the story of a visualization, but distract from the quantitative data~\cite{Bateman.2010}. The goal of our user study is to test low-level tasks such as reading and comparing values correctly, rather than whether aesthetically pleasing designs perform better or worse. That's why we use minimalistic presentations and treat each design in a similar way.

To ensure comparability in the evaluation, we fixed some design criteria and applied them consistently to all five visualizations (\cref{fig:teaser}). The three charts with a color coding -- classic horizon graph, OMH, and OML -- all include a color legend to increase legibility. 
Similair to previous studies on visualizing time-series~\cite{Jabbari.2018, Gogolou.2018, Jabbari.2018b, Perin.2013}, we have chosen not to show the time point labels on the x-axis to avoid semantic meaning that could distract from the actual study tasks.
Y-axis tickmarks are displayed on the main grid lines. Previous research on grid lines shows that, on the one hand, they support the perception of individual values, on the other hand, they can be overwhelming and distracting, which is why they should be used cautiously and not too obtrusively~\cite{Stone.2008, Few.2005, Bartram.2011}. Therefore, we decided to include only one grid line indicating the mantissa value $5$ in addition to the main grid lines used to distinguish the orders of magnitude. The lines within the sections have been drawn with less opacity so that they do not interfere with the visualization.

\begin{figure}
 \centering
 \includegraphics[width=\columnwidth]{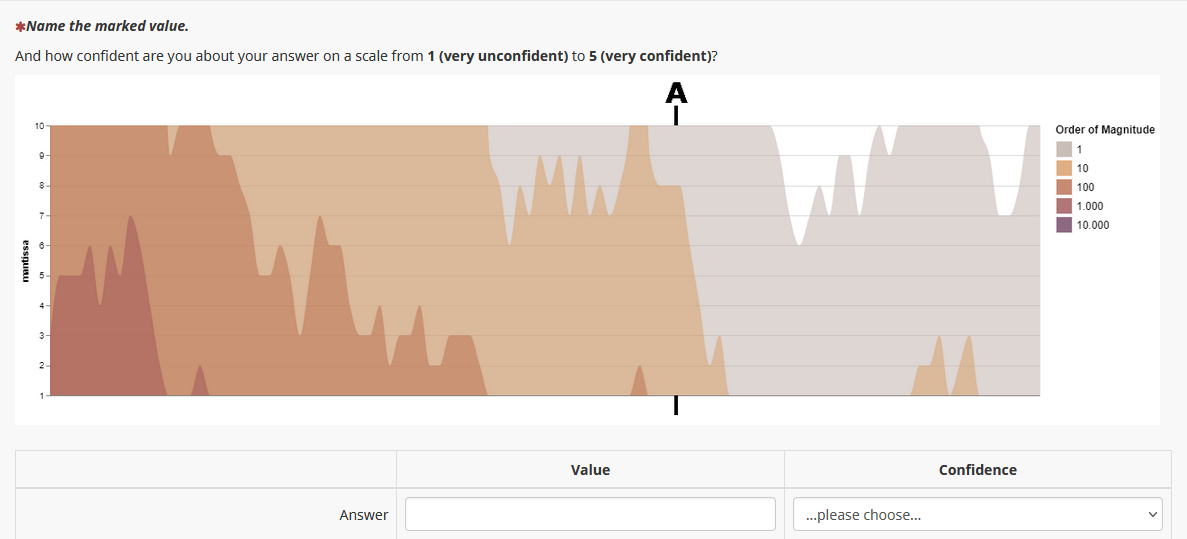}
 \caption{The interface of the user study. This example shows a reading task for the OMH design. The participants type a numerical answer for value \textit{A} in a text box and indicate their confidence level on a drop-down Likert scale.}
 \label{fig:interface}
\end{figure}

An additional element that can interfere with the perception of the visualizations are the visual markers used to highlight the data point of interest in the study tasks. We tried different types of visual markers such as arrows or vertical lines inside the  visualizations, but found that these distracted from the visualization of the data. Therefore, we use a visual marker outside the visualization. To create a comparable environment to the previous horizon graphs studies~\cite{Heer.2009, Jabbari.2018, Jabbari.2018b}, we used the same marker design. The data points are highlighted by two small vertical lines at the top and the bottom of the visualization, as well as a letter above the upper line (see \cref{fig:interface}).

All charts shown in the study had a size of 972x350px to ensure that the visualizations are fully visible on the display for all participants without scrolling. Since we do not consider multiple time-series, each trial contains exactly one visualization.

\subsection{Data}
\label{sec:data}
In comparable studies on large value ranges~\cite{Hlawatsch.2013, Borgo.2014, Hohn.2020}, the data covered a range of $[0, 10000]$ with only integer mantissa. We extend this value range to a maximum value of $100000$, so that our data can be described by $m \cdot 10^{e}$ where $1 \leq m \leq 10$, $0 \leq e \leq 4$ and $m,e \in \vmathbb{Z}$. As we do not focus on streaming applications or long time-series, each data set consists of 100 time steps for consistency.

We used synthetic data sets generated by a constrained random walk. A uniform random value between 1000 and 10000 is the starting point of the walk, from which the mantissa is changed by a uniform random value between $[-2,2]$ at each step. Smoothed generated walks might cover only a fraction of time-series that occur in reality, but they help us to control the study conditions and ensure an equal level of difficulty for all trials. Moreover, they are already used for data generation in previous studies on horizon graphs~\cite{Heer.2009, Jabbari.2018, Jabbari.2018b}.

\subsection{Tasks}
\label{sec:tasks}
In line with state-of-the-art task taxonomies, we focus on four low-level tasks~\cite{Brehmer.2013, Amar.2005, Valiati.2006} corresponding to our hypothesis H1. The selected tasks are a combination of common tasks for both areas we relate to in this paper -- the visualization of time-series data and data with large value ranges. Discrimination and estimation tasks are mostly examined in studies on time-series visualizations~\cite{Heer.2009, Jabbari.2018, Jabbari.2018b, Javed.2010, Perin.2013, Aigner.2012, Kerracher.2015}, while identification and trend analysis tasks can be found in both fields~\cite{Federico.2014, Aigner.2012, Borgo.2014, Hohn.2020, Braun.2022, Hlawatsch.2013, Kerracher.2015}. 

In order to ensure that the trials were comparable and to provide a larger set of answers for the analysis, we split each task into three conditions. The marked values are selected from the previously generated data sets depending on these conditions. We reduced potential bias from the position of these marked values by randomization and multiple trials (three trials per task with different data conditions). The conditions per task we are investigating are as follows:

\paragraph{Identification -- \textit{Value Reading}}
The aim of the "Identification" task is to read individual values as accurately as possible. In each trial, the participant is shown a data set with a marker at a particular value. The target data point is randomly selected with constraints from these conditions:
\begin{itemize}[noitemsep, nolistsep]
\item Condition 1: The value is on a grid line.
\item Condition 2: The value is in a high order of magnitude ($10^{3}$ or $10^{4}$) and not on a grid line. 
\item Condition 3: The value is in a low order of magnitude ($10^{1}$ or $10^{2}$) and not on a grid line.
\end{itemize}
The participants enter their identified value manually via a text box. The interface of an example identification question is shown in \cref{fig:interface}.

\paragraph{Discrimination -- \textit{Value Comparison}}
The aim of the "Discrimination" task is to compare two values and determine which one is larger. As we are not looking at multiple time-series, the discrimination between the two data points takes place in one visualization. In each trial, the participant is shown a data set with two markers at different values. The data points are marked with $A$ and $B$, with $A<B$ on the x-axis. The selection of data points is random, with constraints imposed by the following conditions:
\begin{itemize}[noitemsep, nolistsep]
\item Condition 1: The values are in the same order of magnitude.
\item Condition 2: The values are in neighbouring orders of magnitude.
\item Condition 3: The values are in distinct orders of magnitude with a difference of at least one order of magnitude.
\end{itemize}
The participants select the letter of the larger value from a drop-down menu.

\paragraph{Estimation -- \textit{Difference Determination}}
The estimation task goes one step further with regard to the discrimination task. The aim here is to determine the absolute, quantitative difference between two values in one time-series. Structurally, the task is similar to the discrimination task. Again, the data points are selected randomly according to the same conditions and are also labeled A and B:
\begin{itemize}[noitemsep, nolistsep]
\item Condition 1: The values are in the same order of magnitude.
\item Condition 2: The values are in neighbouring orders of magnitude.
\item Condition 3: The values are in distinct orders of magnitude with a difference of at least one order of magnitude.
\end{itemize}
As in the identification task, the participants enter their answers manually via a text box.

\paragraph{Trend -- \textit{Trend Detection}}
The aim of the "Trend" task is to identify the trend in the visualized data. One visualization without any marker is shown for each trial. For this task, we do not use data generated from a random walk described in \cref{sec:data}. Instead, we created pseudo-random data to simulate specific types of trends. Pseudo-random in this case means that the order of the magnitudes is predetermined to generate a specific kind of trend, but the values for the mantissas are chosen randomly. Although they certainly do not reflect all common trends, we have chosen to test the following three types, based on the selection made in the comparable study task by Höhn et al.~\cite{Hohn.2020}:
\begin{itemize}[noitemsep, nolistsep]
\item Condition 1: Periodic trend.
\item Condition 2: Linear trend.
\item Condition 3: Exponential trend.
\end{itemize}
The participants select the presented trend from a drop-down list containing the options \textit{periodic}, \textit{linear}, \textit{exponential}, and \textit{none}.

\begin{figure*}[tbh!]
    \centering
         \begin{subfigure}[tb]{0.32\textwidth}
             \centering
             \includegraphics[width=\textwidth]{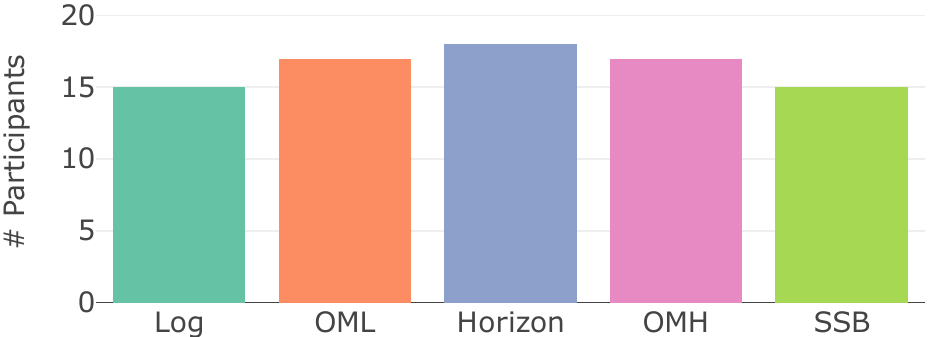}
             \caption{Number of participants per design.}
             \label{fig:design}
         \end{subfigure}
         \hfill
         \begin{subfigure}[tb]{0.32\textwidth}
             \centering
             \includegraphics[width=\textwidth]{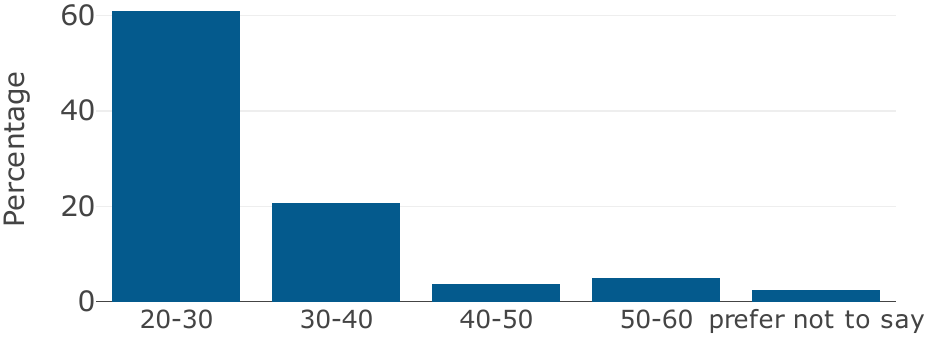}
             \caption{Distribution of participants per age (in \%).}
             \label{fig:age}
         \end{subfigure}
         \hfill
         \begin{subfigure}[tb]{0.32\textwidth}
             \centering
             \includegraphics[width=\textwidth]{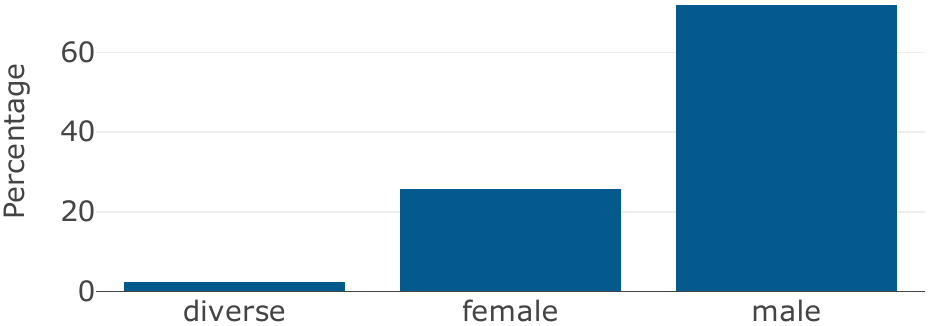}
             \caption{Distribution of participants per gender (in \%).}
             \label{fig:gender}
         \end{subfigure}
            \subfigsCaption{Demographic characteristics of the study participants (excluding participants with color vision deficiencies).}
            \label{fig:demographics}
\end{figure*}

In addition to these tasks, we ask the participants about their confidence in their answers for each trial using a 5-point Likert scale~\cite{Vagias.2006}. We use this as an indicator of the difficulty of interpreting the visualization as perceived by the participants.

\subsection{Experimental Setting}
\label{sec:expsetting}

\paragraph{Procedure}
The study was conducted online and was set up with LimeSurvey~\cite{Limesurvey.2022}. We opted for a between-subject study design~\cite{Charness.2012}, mainly for two reasons: First, it excludes the learning effect that would occur in a within-subject study due to the order in which the designs are processed. Second, the number of trials per participant is reduced, which allowed us to divide the tasks into the different data conditions for more expressive results (see \cref{sec:tasks}).

In the study interface, each page contained one trial and participants had to manually click a button to move to the next page so that we could store the answers given and measure the response time per trial. A back button has been omitted to avoid learning effects. As the study was online, we were unable to control the size and properties of participants' screens, but we recommended a 13" or larger screen. The influence of confounding variables, such as different screen sizes or color vision deficiencies, was reduced as much as possible. Potential uncontrollable external influences such as distracting environmental noise was averaged out due to the large number of participants.

The average processing time of the study was about 15 minutes. Each participant had to complete three trials per task (the three data conditions), resulting in a total of $1[design] \times 4[tasks] \times 3[conditions] = 12$ trials per participant. After each task, the participants were invited to have a short break if needed. A unique data set was generated for each trial (see \cref{sec:data}), so that a total of $12[trials] \times 5[designs] = 60$ different data sets were used in the study.

After demographic questions, which included single-choice questions about the participants' gender, age, degree, and experience with time-series visualizations, and a short training phase to introduce the types of tasks and input options, the participants were randomly assigned to one of the five designs. Before the actual tasks, the visualization designs were explained and various attention questions were included between the tasks. At the end of the study, the participants were asked to give free text feedback. The study documentation is presented in the supplementary material.

\paragraph{Participants}
A total of 90 participants (64 male, 23 female, 2 diverse, and 1 prefer not to say) took part in the study (\cref{fig:gender}). The age was distributed between 20 and 60, but the majority of the participants (81.7\%) were between 20 and 40 years old (\cref{fig:age}). As the OMC color scale is not suitable for color vision deficiencies, we use the Ishihara test for color blindness~\cite{Clark.1924} before beginning the study. 8 participants did not answer the questions correctly, and were hence filtered out from the results. All participants came from a university environment, i.e., they were students or had a higher academic degree, and were recruited through advertising in lectures, mailing-lists, and word of mouth. Thus, they should be familiar with exponential notation.

\begin{figure}
 \centering
 \includegraphics[width=\columnwidth]{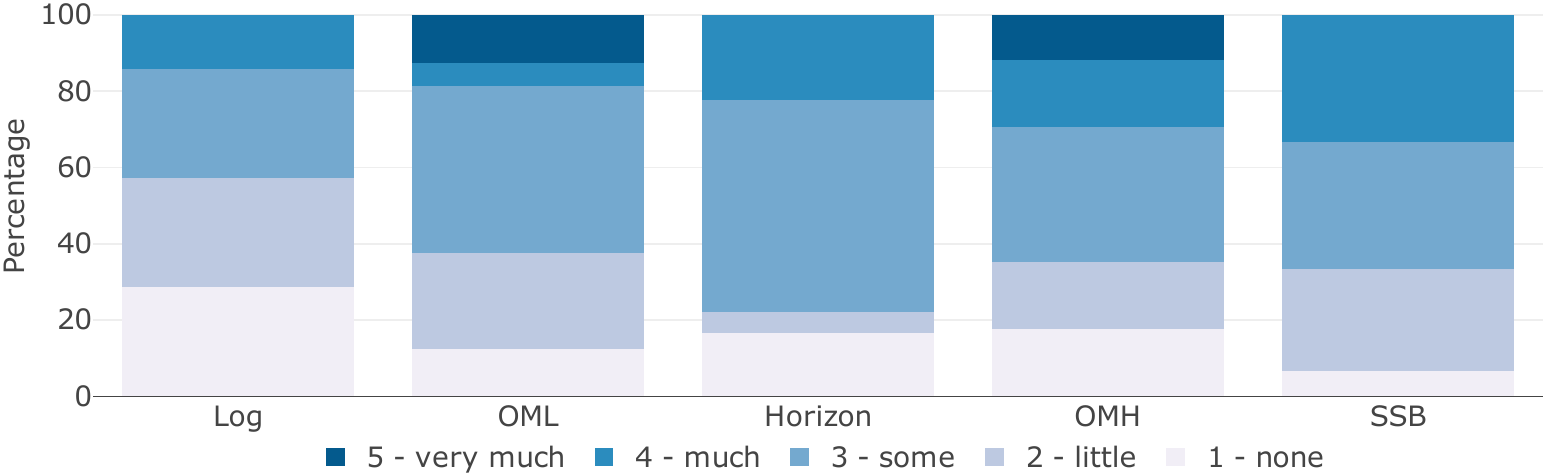}
 \caption{Distribution of participants' self-reported expertise per design.}
 \label{fig:expertise}
\end{figure}

The random allocation of participants among the different designs resulted in the following distribution: 15 people completed the tasks for the log-line chart, 17 for OML, 18 for the classic horizon graph, 17 for OMH, and 15 for SSB (\cref{fig:design}). Due to the random assignment in the between-subject study, there was a potential risk of an expertise bias. We tested for dependence of the participants' reported expertise in time-series using a  $\chi^{2}$-test, which showed no statistical evidence that designs and expertise correlate with each other ($\chi^{2} = 16.168$, $\text{p-value}=0.4431$). Therefore, the results are comparable (see \cref{fig:expertise}).

\section{Results}
\label{sec:results}

In the described study setting, the visualization design as well as the tasks with their different data conditions are the independent variables.
For each task, we analyse error (inaccuracy), confidence, and response time (i.e., the dependent variables of the study), and provide a ranking of the proposed designs based on the results. In addition, we summarize the participants' open-ended feedback.

\begin{figure*}[tbh!]
    \centering
        \begin{subfigure}[b]{0.32\textwidth}
            \centering
            \includegraphics[width=\textwidth]{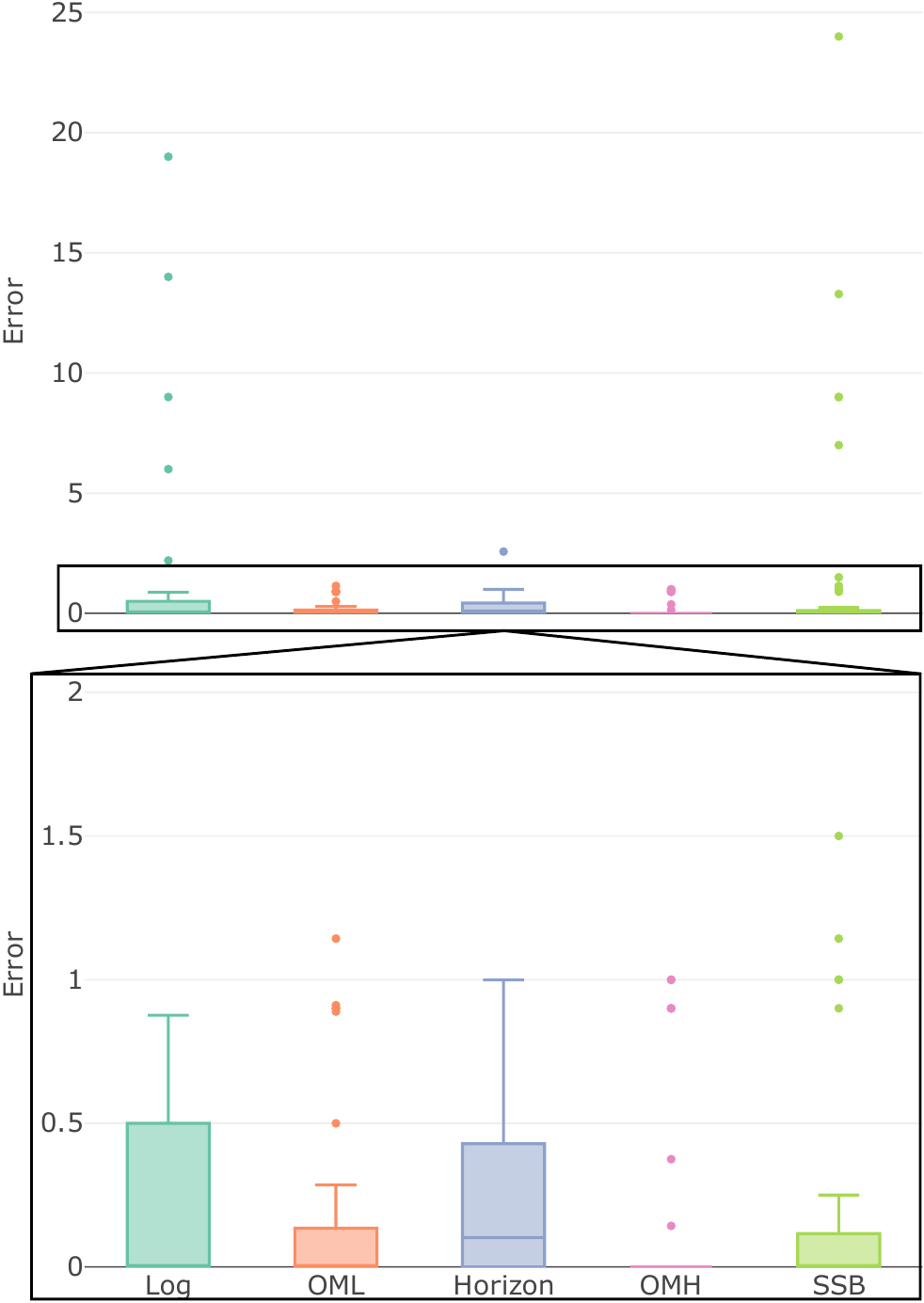}
            \caption{Identification task (bottom: zoomed version).}
            \label{fig:error_identification}
        \end{subfigure}
        \begin{subfigure}[b]{0.32\textwidth}
            \centering
            \includegraphics[width=\textwidth]{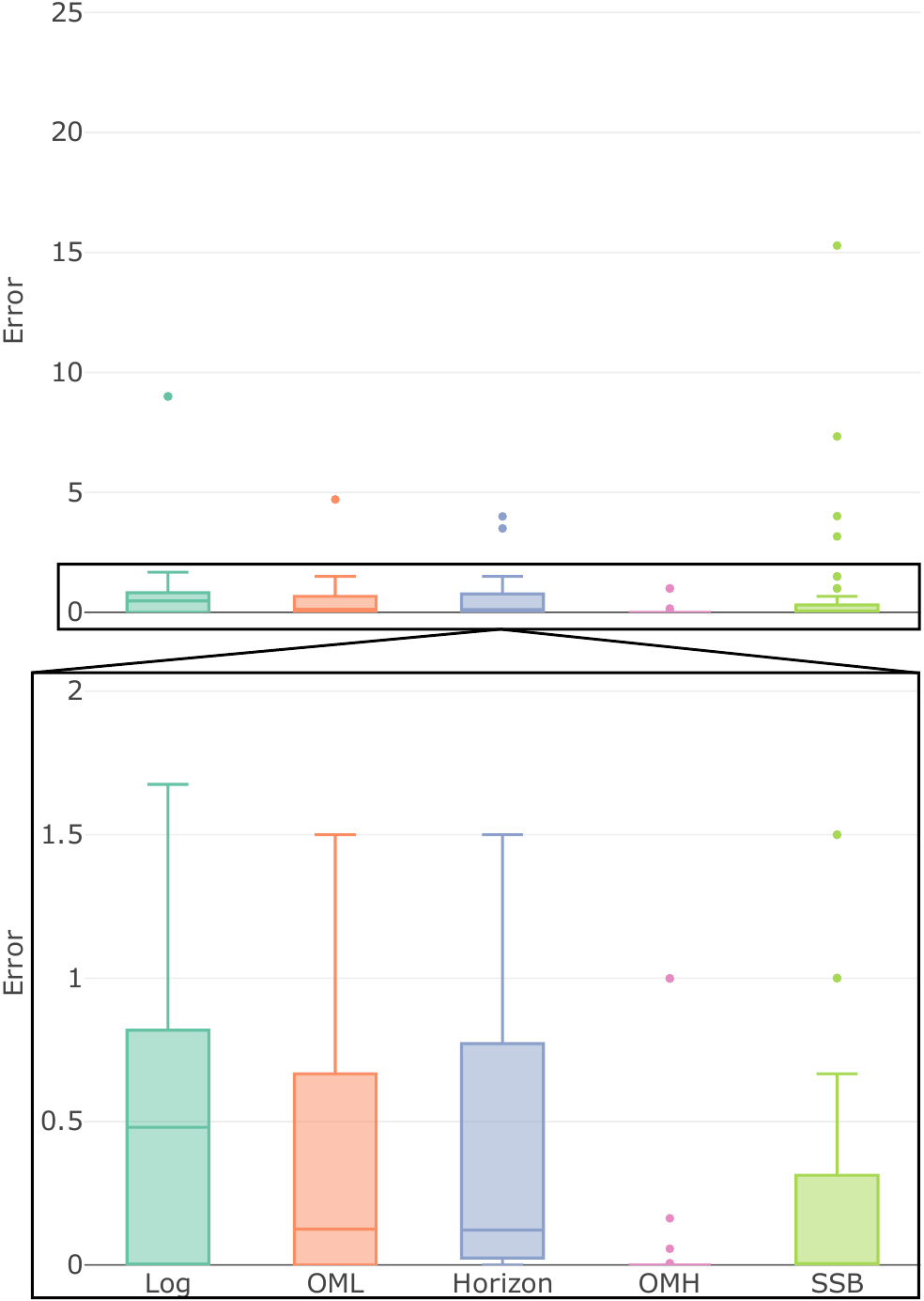}
            \caption{Estimation task (bottom: zoomed version).}
            \label{fig:error_estimation}
        \end{subfigure}
        \begin{subfigure}{0.32\textwidth}
            \begin{subfigure}{\textwidth}
                \centering
                \includegraphics[width=\textwidth]{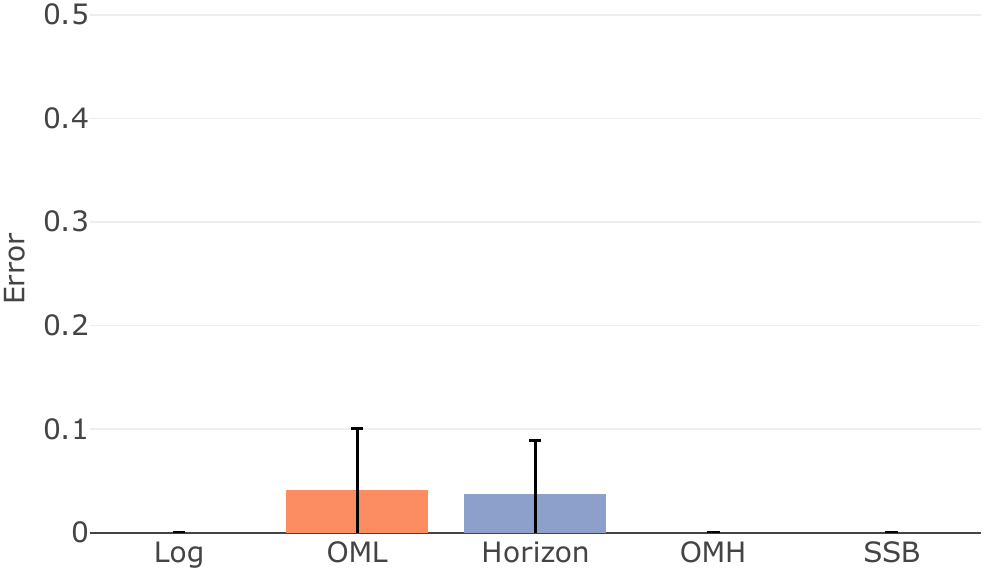}
                \caption{Discrimination task (mean error).}
                \label{fig:error_discrimination}
            \end{subfigure}
            
            \vspace{0.13cm}
            \begin{subfigure}[b]{\textwidth}
                \centering
                \includegraphics[width=\textwidth]{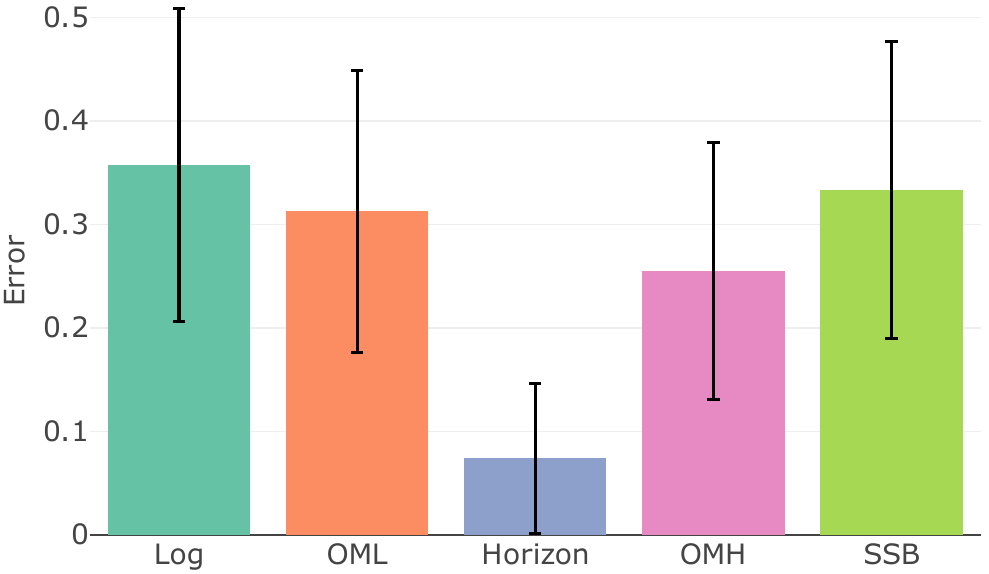}
                \caption{Trend task (mean error).}
                \label{fig:error_trend}
            \end{subfigure}
        \end{subfigure}
    \subfigsCaption{Boxplots of the errors per design for the different study tasks (the lower, the better). }
    \label{fig:errors}
\end{figure*}

\subsection{Analysis}

\paragraph{Error Measurement}
As we have quantitative answers for identification and estimation tasks, but categorical answers for discrimination and trend tasks, we use two different error definitions to test H1.

To allow comparison with the results of Hlawatsch et al.~\cite{Hlawatsch.2013}, we use the same error calculation for the identification and estimation tasks: $\textit{error}=|1-\frac{\textit{response value}}{\textit{correct value}}|$. This measures the relative deviation from the correct value (the lower, the better). This error definition particularly takes into account the characteristics of logarithmic scaled data. Thus, $\textit{error}_{1}=|1-10/100|$ and $\textit{error}_{2}=|1-1000/10000|$ give the same error value ($\textit{error}_{1}=\textit{error}_{2}=0.9$), although the absolute error is different ($90$ resp. $9000$).
Therefore, an error $\geq 1$ indicates an exponent error.

We considered the method of Borgo et al.~\cite{Borgo.2014} to give a 20\% error tolerance to the correct value to evaluate the accuracy of an answer. However, it gives an inaccurate error value due to answers being either just correct or not. They used this method because some of their tasks resulted in uncertainty in the answers. This is not the case for us, as we ask for precisely determinable values in our study. Thus, we decided not to use this method.

For the discrimination and trend tasks, we use the error calculation of Höhn et al.~\cite{Hohn.2020}. Due to the single-choice nature of the questions, there is only the binary result of \textit{true} or \textit{false} in both tasks. By replacing true and false with 1 and 0 (representing 100\% resp. 0\% correctness), we create an accuracy of the answers. Subsequently, these are converted into error values by $1 - \textit{accuracy}$. This transformation preserves that lower error values represent better results.

\paragraph{Significance Tests}
To analyse the results, we used a three-stage significance test for each task. We first ran a Shapiro-Wilk test on the answers and response times to test if the data is normally distributed. The test results showed that there is no normal distribution for any of the tasks. Therefore, we use non-parametric tests for further steps.

The second stage of the analysis is divided into tests for quantitative and categorical data. For the categorical confidence responses, we tested the answers for independence in the designs with the $\chi^{2}$-test. A dependence of confidence and design indicates a significant difference. For the quantitative errors and response times, we used Kruskal-Wallis to test the designs on different means.

As post-hoc analysis for the independent observations we used a Wilcoxon-Mann-Whitney-test for the quantitative answers and a $\chi^{2}$-test for the confidence answers. This was done only for tasks and answers that were found to be significant in the second stage. This allows a comparison of all design combinations and the testing of our hypotheses. The p-values of the combinational comparisons are presented in \cref{sec:results} in the form of triangle matrices.

All test were performed with the standard significance level $\alpha = 0.05$ and a Bonferroni correction factor of 10, corresponding to the number of pairwise comparisons per task and measured aspect.

\subsection{Error Rates}
\begin{figure*}[tbh!]
 \centering
 \includegraphics[width=\textwidth]{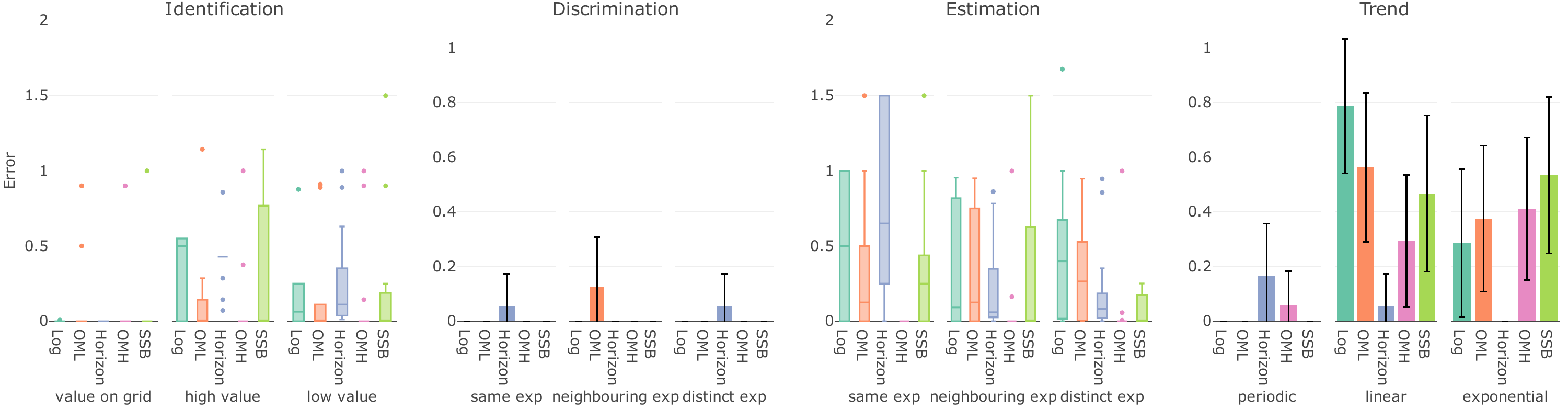}
 \caption{Boxplots of the error rates for the different data conditions per task and design.}
 \label{fig:level}
\end{figure*}

The results of the error analysis are summarized in \cref{fig:errors}. For the identification and estimation tasks, the figures consist of a global box plot zoomed to a range from 0 to 2 in the lower part of the image (\cref{fig:error_identification} and \cref{fig:error_estimation}). Box plots are constructed as follows: The box is bounded by the 25\% and 75\% quantiles of the data and contains a thicker line indicating the median. The whiskers have a maximum length of $1.5 \times \text{[box height]}$, but only extend to the furthest data point within this range. The error bars for the discrimination and trend tasks show the mean error with the 95\% confidence interval (\cref{fig:error_discrimination} and \cref{fig:error_trend}). In \cref{fig:level}, the error rates for the different data conditions are displayed. Averages stated below are adjusted for outliers.

\begin{table}[tb]
    \begin{subtable}[tb]{\columnwidth}
        \centering
        \begin{tabular}{l|c|c|c|c|c}
            \textbf{p-value} & \textit{Log} & \textit{OML} & \textit{Horizon} & \textit{OMH} & \textit{SSB} \\
            \hline
            \textit{Log} & - &  &  &  & \\
            \hline
            \textit{OML} & 1.000 & - &  &  &  \\  
            \hline
            \textit{Horizon} & 1.000 & 0.054 & - &  &  \\
            \hline
            \textit{OMH} & \color{Goldenrod}{0.015} & 0.582 & \color{Goldenrod}{$\approx$0.000} & - &  \\
            \hline
            \textit{SSB} & 1.000 & 1.000 & 0.196 & 0.442 & - \\
        \end{tabular}
    \caption{Identification task.}
    \label{tab:error_ident}
    \end{subtable}
    
    \begin{subtable}[tb]{\columnwidth}
        \centering
        \begin{tabular}{l|c|c|c|c|c}
            \textbf{p-value} & \textit{Log} & \textit{OML} & \textit{Horizon} & \textit{OMH} & \textit{SSB} \\
            \hline
            \textit{Log} & - &  &  &  & \\
            \hline
            \textit{OML} & 1.000 & - &  &  &  \\  
            \hline
            \textit{Horizon} & 1.000 & 1.000 & - &  &  \\
            \hline
            \textit{OMH} & \color{Goldenrod}{$\approx$0.000} & \color{Goldenrod}{$\approx$0.000} & \color{Goldenrod}{$\approx$0.000} & - &  \\
            \hline
            \textit{SSB} & 0.057 & 0.611 & \color{Goldenrod}{0.045} & \color{Goldenrod}{0.004} & - \\
        \end{tabular}
    \caption{Estimation task.}
    \label{tab:error_est}
    \end{subtable}
    
    \begin{subtable}[tb]{\columnwidth}
        \centering
        \begin{tabular}{l|c|c|c|c|c}
            \textbf{p-value} & \textit{Log} & \textit{OML} & \textit{Horizon} & \textit{OMH} & \textit{SSB} \\
            \hline
            \textit{Log} & - &  &  &  & \\
            \hline
            \textit{OML} & 1.000 & - &  &  &  \\  
            \hline
            \textit{Horizon} & \color{Goldenrod}{0.006} & \color{Goldenrod}{0.022} & - &  &  \\
            \hline
            \textit{OMH} & 1.000 & 1.000 & 0.125 & - &  \\
            \hline
            \textit{SSB} & 1.000 & 1.000 & \color{Goldenrod}{0.012} & 1.000 & - \\
        \end{tabular}
    \caption{Trend task.}
    \label{tab:error_trend}
    \end{subtable}
\subfigsCaption{p-values of the pairwise Wilcoxon-test for the error analysis per tasks with significant Kruskal-Wallis-test.}
\label{tab:error}
\end{table}

\paragraph{Identification Task}

Our two new designs OMH and OML performed best for the identification task, while Log and SSB were most susceptible to exponent errors (see \cref{fig:error_identification}). The Kruskal-Wallis-test indicated a significant main effect in error rates ($\chi^{2} = 23.582$, $\text{p-value}=9.686\mathrm{e}{-5}$). The post-hoc pairwise analysis showed (\cref{tab:error_ident}):

\begin{itemize}[noitemsep, nolistsep]
    \item OMH ($\mu = 0$) had a significantly lower error rate than the classic horizon graph ($\mu = 0.21$) and the log-line chart ($\mu = 0.14$).
\end{itemize}

A separate consideration of the error rates for the different data conditions in \cref{fig:level} showed that reading values on grid lines was very easy for all designs, as almost no errors were made. For both the log-line and SSB charts, reading values at higher magnitudes resulted in more errors, while for the classic horizon graph, this was the case at lower magnitudes. Our OMH graph did not lead to increased errors in any of the conditions.

\paragraph{Discrimination Task}
In the discrimination task, only two errors each were made with the OML ($\mu = 0.042$) and classic horizon ($\mu = 0.037$) design (see \cref{fig:error_discrimination}). These very good results indicate that the participants responded to the tasks reasonably. Due to the low number of errors, no significant main effect could be determined with the Kruskal-Wallis-test ($\chi^{2} = 5.5137$, $\text{p-value}=0.2385$). Accordingly, no significant patterns could be observed between the different data conditions (see \cref{fig:level}).

\paragraph{Estimation Task}
OMH performed best in this task. SSB again had the most exponent errors, while the other three designs had similar error rates (see \cref{fig:error_estimation}). The Kruskal-Wallis-test showed a significant main effect ($\chi^{2} = 57.564$, $\text{p-value}=9.422\mathrm{e}{-12}$). The post-hoc pairwise analysis showed (\cref{tab:error_est}):

\begin{itemize}[noitemsep, nolistsep]
    \item OMH ($\mu = 0$) had a significantly lower error rate than all the other designs.
    \item SSB ($\mu = 0.085$) performed significantly better than the classic horizon graph ($\mu = 0.356$).
\end{itemize}

An analysis based on the data conditions showed different effects for the designs (see \cref{fig:level}). With the two line graphs -- log-line and OML --, it was equally difficult to determine differences in values, regardless of the position of the values being compared. The classic horizon graph was less prone to error the further apart the two values were. While most errors in SSB occurred at values in neighbouring orders of magnitude, our OMH design resulted in very low error rates regardless of the conditions.

\paragraph{Trend Task}
For trend detection, the classic horizon graph had the lowest error rate of all designs. An obvious reason for this is the linear scaling of the y-axis exclusively in the classic horizon graph. Of the remaining designs, our new OMH and OML approaches performed best (see \cref{fig:error_trend}). A significant main effect in the error rates was detected by the Kruskal-Wallis-test ($\chi^{2} = 13.711$, $\text{p-value}=0.008$). The post-hoc pairwise analysis showed (\cref{tab:error_trend}):

\begin{itemize}[noitemsep, nolistsep]
    \item The classic horizon graph ($\mu = 0.074$) had a significantly lower error rate than all other designs except the OMH ($\mu = 0.255$).
\end{itemize}

\Cref{fig:level} shows the data conditions. The periodic trend was very well detected by the participants in almost all designs. Note that here most of the errors occurred with the classic horizon graph, which otherwise had the lowest error rates. For the line charts OML and log-line, it was more difficult to detect the linear trend than the exponential trend, while for OMH and SSB, the opposite was true.\\

In sum, the error analysis shows these rankings per task:

\begin{tabular}{ll}
     Identification: & \textit{OMH} $\succeq$ \textit{OML} $\succeq$ SSB $\succeq$ Log $\succeq$ Horizon\\
     Discrimination: & \textit{OMH} $\sim$ Log $\sim$ SSB $\succeq$ Horizon $\succeq$ \textit{OML}\\
     Estimation: & \textit{OMH} $\boldsymbol{\succ}$ SSB $\succeq$ \textit{OML} $\succeq$ Horizon $\succeq$ Log\\
     Trend: & Horizon $\succeq$ \textit{OMH} $\succeq$ \textit{OML} $\succeq$ SSB $\succeq$ Log
\end{tabular}

These results partially confirm our hypothesis \textbf{H1}, although OML ranks in the middle and the standard horizon graph has the lowest error rates in trend detection.

\subsection{Confidence}
\Cref{fig:conf} provides an overview of the Likert-score distributions of the confidence per task and design. The categorical responses were quantified for the means reported below, but for statistical analysis, the designs and categorical answers were tested for independence using the $\chi^{2}$-test. For confidence, the higher the value, the better, as this indicates the participants found the visualization easier to interpret. Hence, it provides an additional qualitative indicator of the subjective trustfulness of the visualizations in addition to the quantitative measures.

\begin{table}[tb]
    \begin{subtable}[tb]{\columnwidth}
        \centering
        \begin{tabular}{l|c|c|c|c|c}
            \textbf{p-value} & \textit{Log} & \textit{OML} & \textit{Horizon} & \textit{OMH} & \textit{SSB} \\
            \hline
            \textit{Log} & - &  &  &  & \\
            \hline
            \textit{OML} & 0.185 & - &  &  &  \\  
            \hline
            \textit{Horizon} & 0.817 & 0.159 & - &  &  \\
            \hline
            \textit{OMH} & 0.122 & 0.567 & \color{Goldenrod}{0.047} & - &  \\
            \hline
            \textit{SSB} & \color{Goldenrod}{0.016} & \color{Goldenrod}{0.006} & \color{Goldenrod}{0.034} & \color{Goldenrod}{$\approx$0.000} & - \\
        \end{tabular}
    \caption{Discrimination task.}
    \label{tab:conf_disc}
    \end{subtable}
    
    \begin{subtable}[tb]{\columnwidth}
        \centering
        \begin{tabular}{l|c|c|c|c|c}
            \textbf{p-value} & \textit{Log} & \textit{OML} & \textit{Horizon} & \textit{OMH} & \textit{SSB} \\
            \hline
            \textit{Log} & - &  &  &  & \\
            \hline
            \textit{OML} & 0.310 & - &  &  &  \\  
            \hline
            \textit{Horizon} & \color{Goldenrod}{0.007} & \color{Goldenrod}{0.007} & - &  &  \\
            \hline
            \textit{OMH} & \color{Goldenrod}{$\approx$0.000} & \color{Goldenrod}{$\approx$0.000} & \color{Goldenrod}{$\approx$0.000} & - &  \\
            \hline
            \textit{SSB} & 0.178 & 0.800 & \color{Goldenrod}{0.004} & \color{Goldenrod}{0.001} & - \\
        \end{tabular}
    \caption{Estimation task.}
    \label{tab:conf_est}
    \end{subtable}
\subfigsCaption{p-values of the pairwise $\chi^{2}$-test for the confidences per tasks with significant $\chi^{2}$-test on all designs.}
\label{tab:conf}
\end{table}

\paragraph{Identification Task}
The usage of our new two designs OMH ($\mu = 4.353$) and OML ($\mu = 4.021$) led to the most confidence in the identification task. The confidences for the follow-up designs Log ($\mu = 3.738$) and SSB ($\mu = 3.733$) were distributed similar, while the classic horizon graph got the least confidence ($\mu = 3.500$). Using the $\chi^{2}$-test, no significant dependence between designs and confidences was found ($\chi^{2} = 25.742$, $\text{p-value}=0.058$).

\paragraph{Discrimination Task}
In general, the participants were very confident in all designs for the discrimination tasks, which correlates to the very few errors made in these tasks. Our new approaches OMH ($\mu = 4.980$) and OML ($\mu = 4.938$) have even received exclusively the answers "5 -- very confident" and "4 -- confident" from the Likert-scale. The confidences in the log-line chart ($\mu = 4.619$) and the classic horizon graph ($\mu = 4.611$) were very close by, while SSB led to the least confidence ($\mu = 4.356$). The $\chi^{2}$-test indicated a significant dependence between design and confidence ($\chi^{2} = 47.342$, $\text{p-value}=6.024\mathrm{e}{-5}$). The post-hoc pairwise analysis showed (\cref{tab:conf_disc}) the following significant dependencies:

\begin{itemize}[noitemsep, nolistsep]
    \item SSB led to significantly lower confidence than all other designs.
    \item Participants were significantly more confident using OMH than the classic horizon graph.
\end{itemize}

\paragraph{Estimation Task}
For the estimation tasks, our OMH graph provided the most confidence ($\mu = 4.235$), followed by the SSB chart ($\mu = 3.356$), the OML ($\mu = 3.313$), and log-line chart ($\mu = 3.048$) charts. Participants had the least confidence in the classic horizon graph ($\mu=2.537$). A significant dependence was shown by the $\chi^{2}$-test ($\chi^{2} = 91.459$, $\text{p-value}=1.346\mathrm{e}{-12}$). The post-hoc pairwise analysis showed (\cref{tab:conf_est})  the following significant dependencies:

\begin{itemize}[noitemsep, nolistsep]
    \item OMH led to significantly higher confidence than all other designs.
    \item Participants were significantly less confident using the classic horizon graph than all other designs.
\end{itemize}

\paragraph{Trend Task}
Confidence was very close for trend detection. Despite having the lowest error rate for this task, the classic horizon graph led just to the second highest confidence ($\mu = 4.222$) behind our OML design ($\mu = 4.396$). The order behind is log-line chart ($\mu = 4.167$), OMH ($\mu = 4.020$), and SSB ($\mu = 3.822$). Due to the similar distributions, the $\chi^{2}$-test showed no significant dependence between designs and confidences ($\chi^{2} = 21.711$, $\text{p-value}=0.153$). \\

Given the results from the confidence analysis, the following design rankings per task are suggested:

\begin{tabular}{ll}
     Identification: & \textit{OMH} $\succeq$ \textit{OML} $\succeq$ Log $\succeq$ SSB $\succeq$ Horizon\\
     Discrimination: & \textit{OMH} $\succeq$ \textit{OML} $\succeq$ Log $\succeq$ Horizon $\boldsymbol{\succ}$ SSB\\
     Estimation: & \textit{OMH} $\boldsymbol{\succ}$ SSB $\succeq$ \textit{OML} $\succeq$ Log $\boldsymbol{\succ}$ Horizon\\
     Trend: & \textit{OML} $\succeq$ Horizon $\succeq$ Log $\succeq$ \textit{OMH} $\succeq$ SSB
\end{tabular}

Hypothesis \textbf{H2} is partially confirmed, since OMH and OML lead the ranking together in almost all tasks.


\subsection{Response Time}
\begin{figure*}[tbh!]
    \centering
         \begin{subfigure}[tb]{\textwidth}
             \centering
             \includegraphics[width=\textwidth]{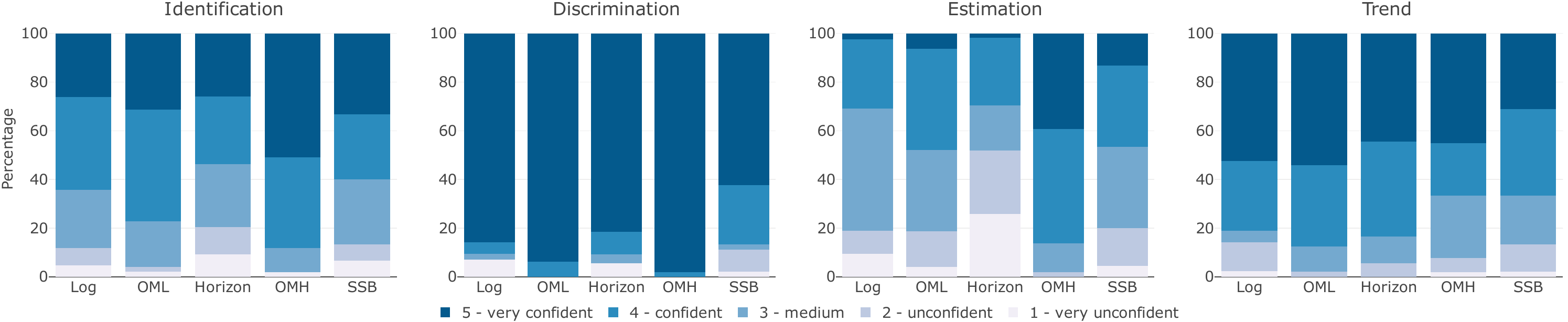}
             \caption{Distribution of the confidence answers per task (the higher the better).}
             \label{fig:conf}
         \end{subfigure}
         
         \begin{subfigure}[tb]{\textwidth}
             \centering
             \includegraphics[width=\textwidth]{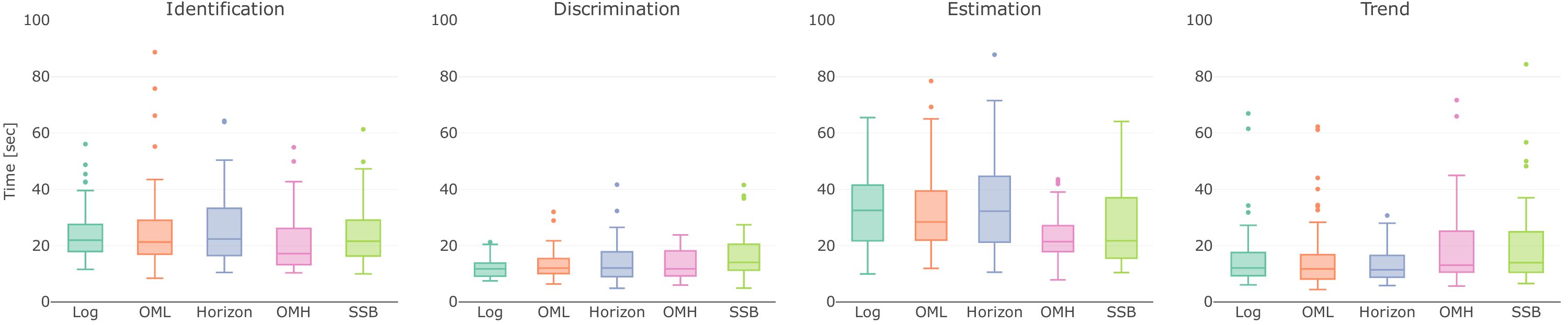}
             \caption{Boxplots of the response times per task in seconds (the lower the better).}
             \label{fig:time}
         \end{subfigure}
            \subfigsCaption{Confidence distributions and response times per task and design.}
            \label{fig:conf_time}
\end{figure*}

A summary of the participants' response times in seconds is shown in \cref{fig:time} using the same box plot design as for the error rates, with lower values being better.

\begin{table}[tb]
    \begin{subtable}[tb]{\columnwidth}
        \centering
        \begin{tabular}{l|c|c|c|c|c}
            \textbf{p-value} & \textit{Log} & \textit{OML} & \textit{Horizon} & \textit{OMH} & \textit{SSB} \\
            \hline
            \textit{Log} & - &  &  &  &  \\
            \hline
            \textit{OML} & 1.000 & - &  &  &  \\  
            \hline
            \textit{Horizon} & 1.000 & 1.000 & - &  &  \\
            \hline
            \textit{OMH} & 1.000 & 1.000 & 1.000 & - &  \\
            \hline
            \textit{SSB} & \color{Goldenrod}{0.019} & 0.207 & 0.578 & 0.242 & - \\
        \end{tabular}
    \caption{Discrimination task.}
    \label{tab:time_disc}
    \end{subtable}
    
    \begin{subtable}[tb]{\columnwidth}
        \centering
        \begin{tabular}{l|c|c|c|c|c}
            \textbf{p-value} & \textit{Log} & \textit{OML} & \textit{Horizon} & \textit{OMH} & \textit{SSB} \\
            \hline
            \textit{Log} & - &  &  &  &  \\
            \hline
            \textit{OML} & 1.000 & - &  &  &  \\  
            \hline
            \textit{Horizon} & 1.000 & 1.000 & - &  &  \\
            \hline
            \textit{OMH} & \color{Goldenrod}{$\approx$0.000} & \color{Goldenrod}{0.004} & \color{Goldenrod}{0.004} & - &  \\
            \hline
            \textit{SSB} & 0.098 & 0.192 & 0.196 & 1.000 & - \\
        \end{tabular}
    \caption{Estimation task.}
    \label{tab:time_est}
    \end{subtable}
\subfigsCaption{p-values of the pairwise Wilcoxon-test for the response time analysis per tasks with significant Kruskal-Wallis-test.}
\label{tab:timer}
\end{table}

\paragraph{Identification Task}
Participants responded fastest with our OMH design for the identification tasks ($\mu = 21.228$).
The Log ($\mu = 24.470$) and SSB ($\mu = 24.641$) had almost equal response times being three seconds slower, while the classic horizon graph ($\mu = 26.183$) and OML ($\mu = 29.689$) were the slowest.
The Kruskal-Wallis-test showed no significant main effect ($\chi^{2} = 7.874$, $\text{p-value}=0.096$). 

\paragraph{Discrimination Task}
The discrimination tasks were overall the tasks with the lowest response times, and all designs have a maximum mean difference of less than five seconds. The log-line chart was the fastest ($\mu = 12.142$), followed by OML ($\mu = 13.299$), OMH ($\mu = 13.414$), and the classic horizon graph ($\mu = 14.131$), while the participants needed the most time for the SSB chart ($\mu = 16.853$). A significant main effect in the error rates was detected by the Kruskal-Wallis-test ($\chi^{2}=9.962$, $\text{p-value}=0.041$). The post-hoc pairwise analysis showed (\cref{tab:time_disc}) only one significant difference:

\begin{itemize}[noitemsep, nolistsep]
    \item The log-line chart had significantly lower response times than the SSB design.
\end{itemize}

\paragraph{Estimation Task}
The estimation tasks took the longest of all four tasks, and the response times varied the most between designs, with a maximum mean difference of more than 10 seconds. Our OMH method was the fastest ($\mu = 22.964$), which is consistent with the error results, as were the second lowest response times of SSB ($\mu = 26.249$). Log ($\mu = 32.819$), OML ($\mu = 33.666$), and Horizon ($\mu = 34.094$) were the order behind it. Due to the high variation in the response times, a significant main effect was indicated by the Kruskal-Wallis-test ($\chi^{2}=23.834$, $\text{p-value}=8.625\mathrm{e}{-5}$). The post-hoc pairwise analysis showed (\cref{tab:time_est}) only the following significant differences:

\begin{itemize}[noitemsep, nolistsep]
    \item Our OMH design was significantly faster than all other designs except SSB.
\end{itemize}

\paragraph{Trend Task}
Matching the error and confidence results, the classic horizon graph got the fastest responses ($\mu = 13.296$). This was followed by the two line graphs Log ($\mu = 15.751$) and OML ($\mu = 16.348$), while the OMH ($\mu = 19.396$) and SSB ($\mu = 20.520$) tasks were the slowest to complete. The Kruskal-Wallis-test showed no significant main effect ($\chi^{2}=8.651$, $\text{p-value}=0.070$). \\

The following design rankings per task are suggested based on the response time analysis:

\begin{tabular}{ll}
    Identification: & \textit{OMH} $\succeq$ Log $\succeq$ SSB $\succeq$ Horizon $\succeq$ \textit{OML}\\
    Discrimination: & Log $\succeq$ \textit{OML} $\succeq$ \textit{OMH} $\succeq$ Horizon $\succeq$ SSB\\
    Estimation: & \textit{OMH} $\succeq$ SSB $\succeq$ Log $\succeq$ \textit{OML} $\succeq$ Horizon\\
    Trend: & Horizon $\succeq$ Log $\succeq$ \textit{OML} $\succeq$ \textit{OMH} $\succeq$ SSB\\
\end{tabular}
With these results, hypothesis \textbf{H3} is rejected, since in almost all cases, OMH and OML are not significantly worse than the others. In fact, there are among the fastest for some tasks.


\subsection{Free Text Feedback}
We received open-ended feedback from 30 participants at the end of the study. From this feedback, the introductions to the tasks and visualization techniques seemed to be understandable for the participants. 

We got most (8) comments on the OMH graph. Participants found our approach to be exciting and mentioned that "using color to present the exponential is a very good visualization method". It was noted that the choice of colors could be improved to make it easier to distinguish between the bands. Using color as a visual channel to make it easier to read accurate values was also mentioned for the OML chart. In general, the responses to both the OMH and OML designs were quite positive. Moreover, the feedback confirmed the results that "trend detection was [the] most difficult" task with the OMH design.

The standard horizon graph received mixed feedback. While it was noted that the design was helpful for discrimination, it was criticized that the overlapping areas with the corresponding value ranges distracted from the actual content of the graph. 

For the log-line graph and the SSB chart participants indicated that these designs were "not suitable for reading concrete data or determining the difference".

\section{Discussion}
\label{sec:discussion}

Order of magnitude horizon (OMH) graphs either outperform significantly or perform comparably well to their currently used counterparts (log-line charts, SSB, and classic horizon graphs) in identification, estimation, and discrimination tasks in terms of error, confidence, and response times. Order of magnitude line (OML) charts, however, do not perform statistically better on error and response times. This confirms the conclusions of previous works~\cite{Braun.2022, Hohn.2020, Borgo.2014, Hlawatsch.2013} from the area of visualizing large value ranges, that the separate representation of mantissa and exponent is suitable for large value ranges (\textbf{H1}). Moreover, it appears that this separation needs to be \textit{discrete}, as the continuous separation in the case of OML does not seem to be sufficient. For trend detection, the classic horizon graph provides significant lower error rates compared to all designs but OMH (lower, but not significant). Horizon graph's significantly better performance is an indication that linear axis scaling is more appropriate for this type of task. It is interesting to note the comparable performance of the two horizon-based designs, with OMH exhibiting signs of speed-accuracy trade-off (SAT) effects (even if not significant) which may indicate participants trading speed for accuracy while undertaking a complex task with a new design. An interesting line of further investigation would be to examine if increased familiarity and usage of OMH could potentially reduce this apparent trade-off and increase performance similarity. These could indicate an inherent effect of the design itself which leverages perceptual grouping.

In general, however, the results of the individual designs seem to be dependent on the values of the queried data points and their ratios, as the analysis of the different data conditions shows. 
For example, the SSB seems to be particularly difficult to read in high orders of magnitude, despite its design being intended to favor large value ranges. Furthermore, the exponential trend is easier to detect than the linear trend in the two line charts (log-line and OML), whereas the opposite is true for OMH and SSB. In summary, further guidelines for using different designs in specific data and task conditions may emerge.

Another interesting aspect is the advantage of using color as a visual variable. The large amount of errors in the exponent of the log-line and SSB charts in identification and estimation tasks suggests that color coding supports the detection of the correct exponents, as these are the only two designs not using color as a visual variable. An additional indicator for this is that the participants had the highest confidence with either OMH or OML (which use color coding) in all tasks (\textbf{H2}). Since OMH and the classic horizon graph both use the design principles of the two-tone pseudo coloring~\cite{Saito.2005}, the improved performance of OMH is most likely due to the novel y-axis composition.

Our OML technique uses the order of magnitude colors (OMC) colormap of Braun et al.~\cite{Braun.2022}, and our results support the usefulness of their approach as it leads to improved error rates compared to the standard log-line chart. The OMC color favors ``banding effects'' when crossing magnitude thresholds, which seems to have contributed to improve the overall visual clarity in our OML layout, although we have not formally tested this in our study. The improved error rates suggests OML to be worthy of further investigation in those domains where not only large value ranges are present, but quantitative discrimination of these magnitudes is a task of major importance. In addition, as pointed out by Borkin et al.~\cite{Borkin.2013}, the low data-to-ink ratio in OMH and OML is an indication that these two visualizations may be more memorable than standard visualization techniques.

Participants had no previous experience with our new visual layouts, and we carefully designed our studies to minimize possible learning effects. We were therefore positively surprised by the fact that no significant global trade-off effect in the response times, confidence, and error rates was detected. Only (not significant) SAT symptoms were detected for OML in the Identification task, and OMH in the Trend detection tasks (as previously discussed), with participant trading speed for accuracy in both cases. This seems to suggest that the cognitive load introduced by the new design was negligible (\textbf{H3}).
The consistency of the subjective confidence responses with the performance of the designs implies that our new approaches were correctly interpreted. It would be interesting to explore further how easily our novel design can be learned when applied to more complex scenarios, as also suggested in~\cite{Zhao.2017}.

\section{Limitations and Future Work}
\label{sec:limitations}

There are still some limitations to our visualization techniques that subsequently provide opportunities for future work.

As the error rates analysis showed and the participant's feedback confirmed, our designs are not well suited for trend detection. It would be interesting to see if an approach can be developed that works well for trend tasks and one or more of the other tasks.
Despite the great results of our OMH approach, participants' comments also indicate that there is room for improvement in the choice of colors for the different orders of magnitude.
The analysis also showed slower response times of OML charts compared to log-line charts. This could be either due to the novelty of OML charts, or the increased cognitive load from double encoding the values by position and color. Although the free feedback suggests the former, this cannot be proven by our study and requires further investigation. 

Although a participant pool from university is not representative for the entire society, it made sure that everyone knew the exponential notation.
We are aware that our study covers only part of the possible types of tasks and data, and that it would be interesting to investigate different types of both (e.g. multiple time-series). Nevertheless, it was sufficient to test our hypotheses. In particular, value scales with both positive and negative values have not been considered in this paper. As mentioned in \cref{sec:OMH}, this can be relatively easily implemented for the OMH design. For the OML chart, on the other hand, further considerations are necessary.

Our approach leverages the effectiveness of visual encoding rather than relying on user interaction (e.g., zooming). The ability to visualize the full data range reduces the need for direct manipulation of the visualization. This has the advantage of being able to maintain context and ensure consistency across different analytical tasks, as well as being more suitable for print. However, it may be possible to integrate and adapt interaction techniques to further increase the effectiveness (especially for long time-series) of the visual encoding in future work.

Finally, our work has focused on orders of magnitude with base 10, as these are the most common. Base 2 and base $e$ would require more colors due to the high exponent range, and vice versa for larger bases. 

\section{Conclusion}
\label{sec:conclusion}
Our work presents two novel visualization designs to support the display of large value ranges in time-series data: The order of magnitude horizon chart and the order of magnitude line chart. Using an empirical user study, we showed the advantages of our designs in accuracy, confidence, and response time for identification, discrimination, and estimation tasks. The performance of our visualization designs for large value ranges is partially dependent on the order of magnitude of the values to be evaluated, with values between magnitudes being harder to compare. Our results confirm previous works in the field of visualization of large value ranges~\cite{Braun.2022, Hohn.2020, Borgo.2014, Hlawatsch.2013} that the separation of the mantissa and exponent is beneficial for the perception of values from large value ranges.


\section*{Supplemental Materials}
\label{sec:supplemental_materials}

Supplemental material includes: 1) Python code (Jupyter-notebook) for the data and design generation; 2) User study documentation including the stimuli as a PDF file; and 3) Anonymized study results as a CSV file.

\acknowledgments{%
  The authors would like to thank all study participants for taking part in the evaluation and the reviewers, whose suggestions helped to improve this paper. This work has been partially supported by BMBF WarmWorld Project and KPA Intelligent Methods for Earth System Sciences.
}

\bibliographystyle{abbrv-doi-hyperref}

\bibliography{literature}










\end{document}